\documentclass{article}
\usepackage[x11names]{xcolor}
\usepackage{listings}
\usepackage{cite}
\usepackage{float}
\usepackage{adjustbox}

\usepackage{url}
\usepackage[section]{placeins}
\usepackage{amsmath,amssymb,amsfonts}
\usepackage{algorithmic}
\usepackage{multirow}
\usepackage{booktabs}
\usepackage{caption}
\usepackage{subcaption}
\usepackage{graphicx}
\usepackage{hyperref}
\hypersetup{
colorlinks=true,
linkcolor=red,
anchorcolor=black,
citecolor=blue,
filecolor=cyan,
menucolor=red,
runcolor=cyan,
urlcolor=magenta
}
\usepackage{subcaption}
\usepackage{textcomp}
\usepackage{xcolor}
\usepackage{geometry}
\usepackage[utf8]{inputenc}
\usepackage{enumerate}
\makeatletter
\newcommand*{\rom}[1]{\expandafter@slowromancap\romannumeral #1@}
\makeatother
\usepackage{times}
\usepackage{alphabeta}
\usepackage{enumitem}
\usepackage{fontawesome}

\title{\textbf{System Calls for Malware Detection and Classification: Methodologies and Applications}}
\author{
{Bishwajit Prasad Gond and Durga Prasad Mohapatra} \\
\textit{National Institute of Technology Rourkela, Odisha, India.}\\
{
ORCID iD: \href{https://orcid.org/0000-0003-3640-0463}{0000-0003-3640-0463}, \href{https://orcid.org/0000-0002-4824-7091}{0000-0002-4824-7091}
}
\date{}
}

\begin{document}
\maketitle
\section*{Abstract}
As malware continues to become more complex and harder to detect, Malware Analysis needs to continue to evolve to stay one step ahead. One promising key area approach focuses on using system calls and API Calls, the core communication between user applications and the operating system and their kernels. These calls provide valuable insight into how software or programs behaves, making them an useful tool for spotting suspicious or harmful activity of programs and software. This chapter takes a deep down look at how system calls are used in malware detection and classification, covering techniques like static and dynamic analysis, as well as sandboxing. By combining these methods with advanced techniques like machine learning, statistical analysis, and anomaly detection, researchers can analyze system call patterns to tell the difference between normal and malicious behavior. The chapter also explores how these techniques are applied across different systems, including Windows, Linux, and Android, while also looking at the ways sophisticated malware tries to evade detection. 
% By weighing the pros and cons of each method, the discussion shows how flexible and scalable system call-based detection can be—making it a strong line of defense against the constantly changing landscape of digital threats.

\section{Introduction}  
Malware ranging from viruses, spyware and worms to trojans, ransomware, and other harmful software and it continues to be a major threat to individuals, businesses, and even the national infrastructure. Traditional methods relying on known static signatures can still be effective in detecting familiar threats, but they often fail when faced with increasingly complex and shape-shifting malware. This has led to a growing focus on behaviour based detection and classification, where system calls have become a central point of interest. Because system calls reflect the core actions a program takes, such as accessing files, managing processes, or connecting to networks, they provide valuable clues about what software is really doing behind the scenes.

In this chapter, we explore how system calls or API calls are used to detect and classify malware, offering a detailed look at both the theory and the real-world application of these techniques. We examine how analyzing system call patterns, both statically and during run-time, can uncover malicious activity, even when malware is hidden or brand new. We also look at how techniques like machine learning and statistical analysis can enhance accuracy and minimize false alarms. By discussing the strengths, limitations, and practical impact of these methods.

The remaining sections of the chapter are formatted as follows: Section \ref{sec:basicconcepts} outlines the basic concepts related to Malware and System Calls. Section \ref{sec:method} details the methodology used to extract System or API Calls. Section \ref{sec:appli} presents the Use cases and their analysis. Finally, Section \ref{sec:future} concludes with future research directions.

\section{Basic Concepts}\label{sec:basicconcepts} 
To effectively detect and classify malware, it is important to understand the basics of system calls, which act as the communication link between regular applications and the core of the operating system i.e the kernel. Simply put, a system call is when a program asks the operating system to carry out a task it can not do on its own, like accessing a file, creating a new process, or sending data over a network. By keeping an eye on these interactions, security tools can get a clearer picture of what a program or software is actually doing and spot unusual behavior that may signal something malicious.

This section lays out the key concepts behind system calls and explains why they play such a crucial role in identifying and classifying malware. It also takes a closer look at how system calls are implemented across different platforms, highlighting the unique challenges and opportunities that come with each environment.

\subsection{Malwares}\label{sec2:malware}
Malware is a short form of ``malicious software" and it is a broad term that refers to any software or code designed with the intent to infiltrate, damage, disrupt, or gain unauthorized access to computer systems, networks, or data. It is typically created with harmful or criminal objectives in mind. This category includes a wide variety of malicious programs, each with its own unique characteristics and purposes to execute.

\subsubsection{Varieties of Malware}
\begin{enumerate}[label=\textbf{\arabic*.}]
    \item \textbf{Viruses:}  Viruses are harmful software that attach to valid executable files or documents. Once the infected file is opened, the virus code is triggered, allowing it to replicate and spread to other files and systems. These malicious programs can cause substantial harm to a computer, such as corrupting or destroying files and applications. A well-known example is the \textbf{ILOVEYOU} virus, which propagated through email in 2000, infecting millions of computers by overwriting files. It was spread through email messages with the subject "ILOVEYOU" and an attached file disguised as a text file (LOVE-LETTER-FOR-YOU.TXT.vbs). 
    
    \item \textbf{Worms:} Worms are self-replicating malware that can spread on their own, without needing a host file. They often take advantage of weaknesses in network protocols or software to move across a network. Worms can overwhelm networks and slow down system performance. A notable example is the \textbf{Conficker} worm, which began infecting millions of computers globally in 2008 by exploiting vulnerabilities in Windows.
    
    \item \textbf{Trojans (Trojan Horses):} Trojans are deceptive malware that masquerades as legitimate software or files. When users unwittingly download and execute them, Trojans can provide attackers with unauthorized access to the victim's system, enabling the theft of sensitive data or the execution of other malicious activities. A notable example is \textbf{Zeus}, a Trojan used to steal banking credentials since 2007.
    
    \item \textbf{Ransomware:} Ransomware encrypts a victim's files or entire system, rendering them inaccessible. Attackers demand a ransom in exchange for the decryption key. It is discouraged to pay the ransom, as it offers no guarantee of data recovery and only serves to incentivize cybercriminals. A prominent example is \textbf{WannaCry}, which in 2017 encrypted data on thousands of systems globally, demanding Bitcoin ransoms.
    
    \item \textbf{Spyware:} Spyware is designed to clandestinely gather information from a victim's computer or device, including keystrokes, browsing history, and personal data. This pilfered information is then transmitted to a remote server, often without the user's awareness or consent. A famous example is \textbf{DarkTequila} spyware, which targeted Latin American users to steal financial data.
    
    \item \textbf{Adware:} Adware is software that shows users unwanted advertisements, often as pop-up windows or banners. While not inherently harmful, adware can be bothersome and may lead to performance issues. Some adware may also collect user data without consent. A common example is \textbf{Fireball}, which in 2017 turned browsers into ad-displaying tools on over 250 million devices.
    
    \item \textbf{Downloader:} A ``downloader'' is a type of malware designed to download and install other malicious software onto a victim's computer or device. Downloaders are often used as a first-stage payload in a cyberattack to deliver more advanced and harmful malware, such as viruses, Trojans, ransomware, or spyware. The primary function of a downloader is to act as a conduit for the delivery of additional malicious code. A well-known example is \textbf{Emotet}, initially a banking Trojan but often used as a downloader for other malware since 2014.
\end{enumerate}

\subsection{Classification and detection of Malwares}\label{sec2:classificationofmalware}
Malware detection and prevention techniques are essential to safeguarding computer systems and data against these various types of threats. This includes using antivirus software, keeping software and systems up-to-date, practising safe online behaviour, and implementing network security measures.

Classifying malware is a critical task in cybersecurity. One effective approach involves analyzing the sequence of Application Programming Interface (API) calls made by a program. API sequence data can provide valuable insights into the behaviour of a program, allowing us to distinguish between benign and malicious software.

% \section{API Sequence Data}

API sequence data is a series of API calls made by a program during its execution. Each API call represents an interaction with the operating system or external libraries. For malware classification, we can collect API sequence data by monitoring a program's execution and recording the order of API calls.

% \section{Classification Method}

We can use machine learning algorithms to classify malware based on API sequence data. One common approach is to extract features from the API sequences and train a classifier to distinguish between benign and malicious behaviour.

\subsection{System Calls for Malware Detection and Classification}
A system call is a programmatic request made by a user-level application to the operating system's kernel to perform privileged operations that the application cannot execute directly. These operations include tasks such as file access (e.g., \texttt{open}, \texttt{write}), process management (e.g., \texttt{fork}, \texttt{execve}), and network communication (e.g., \texttt{socket}, \texttt{connect}). System calls serve as the primary interface between user applications and the operating system, enabling controlled access to hardware and system resources. For example, on Linux, the \texttt{open} system call allows a program to access a file, while on Windows, \texttt{NtCreateFile} serves a similar purpose through the Native API (NTAPI). In the context of malware detection, system calls are critical because they reveal a program's runtime behavior, allowing analysts to identify malicious activities, such as unauthorized file modifications or network data exfiltration, by monitoring sequences and parameters of these calls.

System calls are vital in malware analysis as they provide a low-level view of a program's interactions with the operating system kernel, revealing behaviors critical for detecting malicious activities. These interactions, such as file operations (\texttt{open}, \texttt{NtCreateFile}), process management (\texttt{fork}), or network communications (\texttt{socket}), are monitored using tools like \texttt{strace}, Process Monitor, or Cuckoo Sandbox to identify suspicious patterns, such as file encryption by ransomware or data exfiltration by spyware. Their universal role across platforms (Windows, Linux, Android) ensures consistent analysis, while their inevitability makes them resistant to obfuscation techniques like code packing.

System calls vary across operating systems due to differences in kernel design, architecture, and intended use cases. Below, we discuss how system calls are utilized for malware detection on Windows, Linux, and Android, providing examples for each platform, analyzing their differences, and evaluating their traceability and robustness against issues like process threads, stack overflows, or memory vulnerabilities.

\subsubsection{Windows}
On Windows, system calls are managed through the Native API (NTAPI), a low-level interface exposed by the \texttt{ntdll.dll} library. These calls, prefixed with \texttt{Nt} or \texttt{Zw}, include operations like \texttt{NtCreateFile} (file creation) and \texttt{NtWriteVirtualMemory} (memory manipulation). Malware detection on Windows often involves monitoring these calls to identify suspicious patterns, such as unauthorized file modifications or process injection and memory overflow. Below are the some hooked API call categores and examples.

\subsubsection*{Hooked APIs and Categories}\label{subsec:hookapi}
%%%%%%%%%%%%%%%%%%%%%%%%%%%%%%%%%%%%%%%%%
\subsubsection*{\_\_Notification\_\_}
\_\_missing\_\_, 
\_\_exception\_\_, 
\_\_Process\_\_, 
\_\_Anomaly\_\_ etc
\subsubsection*{Exception}
SetUnhandledExceptionFilter, 
RtlDispatchException, 
RtlAddVectoredExceptionHandler, 
RtlRemoveVectoredExceptionHandler, 
RtlRemoveVectoredContinueHandler etc

\subsubsection*{Certificate}
CertOpenSystemStoreW, CertOpenSystemStoreA, CertOpenStore, CertControlStore, CertCreateCertificateContext, etc

\subsubsection*{Crypto}
Ssl3GenerateKeyMaterial, PRF, CryptUnprotectMemory, CryptUnprotectData, CryptProtectMemory, CryptProtectData, CryptHashMessage, CryptEncryptMessage, etc

\subsubsection*{File}
WSASocketW, WSASocketA, WSASendTo, WSASend, WSARecvFrom, WSARecv, WSAConnect, WSAAccept, TransmitFile, NtQueryInformationFile, NtQueryFullAttributesFile, NtQueryDirectoryFile, NtQueryAttributesFile, NtOpenFile, etc

\subsubsection*{Misc}
WriteConsoleW, WriteConsoleA, GetTimeZoneInformation, GetDiskFreeSpaceW, GetDiskFreeSpaceExW, UuidCreate, GetComputerNameW, GetComputerNameA, LookupAccountSidW, GetUserNameW, GetUserNameA, etc

\subsubsection*{NetAPI}
NetUserGetLocalGroups, NetUserGetInfo, NetShareEnum, NetGetJoinInformation, etc

\subsubsection*{OLE}
OleInitialize, CoInitializeEx, CoCreateInstance, etc

\subsubsection*{Network}
URLDownloadToFileW, ObtainUserAgentString, ConnectEx, GetInterfaceInfo, GetBestInterfaceEx, GetAdaptersInfo, GetAdaptersAddresses, DnsQuery\_W, DnsQuery\_UTF8, DnsQuery\_A, InternetWriteFile, InternetSetStatusCallback, InternetSetOptionA, InternetReadFile, InternetOpenW, InternetOpenUrlW, InternetOpenUrlA, InternetOpenA, InternetGetConnectedStateExW, InternetGetConnectedStateExA, InternetGetConnectedState, InternetCrackUrlW, InternetCrackUrlA, InternetConnectW, InternetConnectA, InternetCloseHandle, HttpSendRequestW, HttpSendRequestA, HttpQueryInfoA, HttpOpenRequestW, HttpOpenRequestA, DeleteUrlCacheEntryW, DeleteUrlCacheEntryA, etc

\subsubsection*{System}
ShellExecuteExW, RtlCreateUserThread, RtlCreateUserProcess, NtCreateThread, NtTerminateThread, NtTerminateProcess, NtSuspendThread, NtSetContextThread, NtResumeThread, NtReadVirtualMemory, NtQueueApcThread, NtOpenThread, NtProtectVirtualMemory, NtOpenSection, NtOpenProcess, NtMapViewOfSection, NtMakeTemporaryObject, NtMakePermanentObject, NtGetContextThread, NtFreeVirtualMemory, NtCreateUserProcess, NtCreateThreadEx, NtCreateSection, NtCreateProcessEx, NtCreateProcess, NtAllocateVirtualMemory, UnhookWindowsHookEx, SetWindowsHookExW, SetWindowsHookExA, SendNotifyMessageW, SendNotifyMessageA, GetKeyboardState, GetKeyState, GetAsyncKeyState, ExitWindowsEx, RtlDecompressFragment, RtlDecompressBuffer, RtlCompressBuffer, NtUnloadDriver, NtLoadDriver, NtDuplicateObject, NtClose, etc

\subsubsection*{Resource}
SizeofResource, LoadResource, FindResourceW, FindResourceExW, FindResourceExA, FindResourceA, etc

\subsubsection*{Services}
StartServiceW, StartServiceA, OpenServiceW, OpenServiceA, OpenSCManagerW, OpenSCManagerA, EnumServicesStatusW, EnumServicesStatusA, DeleteService, CreateServiceW, CreateServiceA, ControlService, etc

\subsubsection*{Synchronization}
timeGetTime, NtQuerySystemTime, NtDelayExecution, NtCreateMutant, GetSystemTimeAsFileTime, GetSystemTime, GetLocalTime, etc

\subsubsection*{UI}
MessageBoxTimeoutW, MessageBoxTimeoutA, LoadStringW, LoadStringA, GetForegroundWindow, FindWindowW, FindWindowExW, FindWindowExA, FindWindowA, DrawTextExW, DrawTextExA, etc\\

\noindent\textbf{Example:} A ransomware might invoke \texttt{NtCreateFile} to access a user documents, followed by \texttt{NtWriteFile} to encrypt the contents. By tracing these calls using tools like Process Monitor or a custom kernel driver, analysts can flag the behavior as malicious.

\subsubsection{Linux}
Linux employs a POSIX-compliant system call interface, accessible via the \texttt{syscall} instruction or higher-level wrappers like \texttt{glibc}. Common system calls include \texttt{open} (file access), \texttt{fork} (process creation), and \texttt{socket} (network operations). Malware detection on Linux leverages tools like \texttt{strace} or \texttt{ptrace} to monitor these calls and detect anomalies, such as unexpected network connections or privilege escalation attempts.

\textbf{Example:} A rootkit might use \texttt{open} to access \texttt{/etc/passwd} and \texttt{write} to modify it stealthily. By analyzing the sequence and parameters of these calls, detection systems can identify the malicious intent.

\subsubsection{Android}
Android, built on a modified Linux kernel, uses system calls inherited from Linux but extends them with a Java-based framework via the Android Runtime (ART). Key system calls include \texttt{read}, \texttt{write}, and \texttt{bind} for I/O and networking, often invoked indirectly through Android APIs. Malware detection on Android typically involves monitoring both native system calls and higher-level API calls using tools like Frida or Xposed.

\textbf{Example:} A spyware app might call \texttt{socket} and \texttt{write} to exfiltrate user data over a network. By hooking these calls at the native level or inspecting API usage, detection systems can pinpoint the malicious activity.

\subsubsection{Differences Between System Calls}
The system calls of Windows, Linux, and Android differ significantly in their implementation and accessibility. Below, we discuss some of the differences:
\begin{itemize}
    \item \textbf{Windows:} Uses a proprietary NTAPI with undocumented calls, making it complex but tightly integrated with the kernel. Calls are invoked via interrupt \texttt{0x2E} or \texttt{SYSENTER}.
    \item \textbf{Linux:} Offers an open, well-documented POSIX interface, invoked via \texttt{syscall} or software interrupts (\texttt{int 0x80}). It is simpler and more transparent than Windows.
    \item \textbf{Android:} Combines Linux system calls with a layered architecture (Java APIs over native calls), adding complexity due to the abstraction but enabling cross-layer analysis.
\end{itemize}

\subsubsection{Ease of Tracing and Robustness}
\begin{itemize}
    \item \textbf{Ease of Tracing:} Linux is the easiest O.S. to trace due to its open-source nature and tools like \texttt{strace}, which provide detailed call logs. Windows tracing is more challenging due to proprietary APIs and anti-debugging techniques, requiring specialized tools like Sysinternals. Android falls in between, as tracing native calls is feasible, but Java-layer obfuscation complicates analysis.
    \item \textbf{Robustness:} Windows is relatively robust against stack or memory overflows in system call handling due to its strict kernel-user separation and memory protection mechanisms. Linux is moderately robust but vulnerable to kernel exploits if improperly configured. Android, inheriting Linux’s kernel, is less robust due to its mobile-optimized design and frequent use of native code, increasing the risk of memory overflows or thread-related vulnerabilities.
\end{itemize}

In summary, Linux offers the most traceable system calls, while Windows provides greater robustness. Android’s hybrid architecture presents unique challenges and opportunities for malware detection, balancing traceability and resilience based on the layer analyzed.

\section{Methodologies for Extracting System Calls}\label{sec:method}
To effectively utilize system calls for malware detection, various methodologies are employed to extract and analyze them. These approaches can be broadly categorized into three categories such as static analysis, dynamic analysis, and sandboxing. Each method offers distinct advantages and challenges, depending on the target platform and the nature of the malware. This section explores these techniques, with a focus on their application to malware detection, and provides an overview of sandboxing tools, including open-source and license/enterprise solutions.

\subsection{Static System Call Analysis}
Static analysis involves examining a program’s code or binary without executing it, aiming to identify system calls and infer potential behavior. This method typically uses disassemblers (e.g., IDA Pro) or decompilers to extract system call references from the executable. For instance, on Windows, static analysis might reveal calls to \texttt{NtCreateProcess} or \texttt{NtOpenKey}, suggesting process creation or registry manipulation. On Linux, tools like \texttt{objdump} can identify calls to \texttt{execve} or \texttt{connect}, indicating process execution or network activity.

\begin{figure}[h]
    \centering
    \begin{subfigure}[b]{0.45\textwidth}
        \centering
        \includegraphics[angle=0,origin=c, width=6cm]{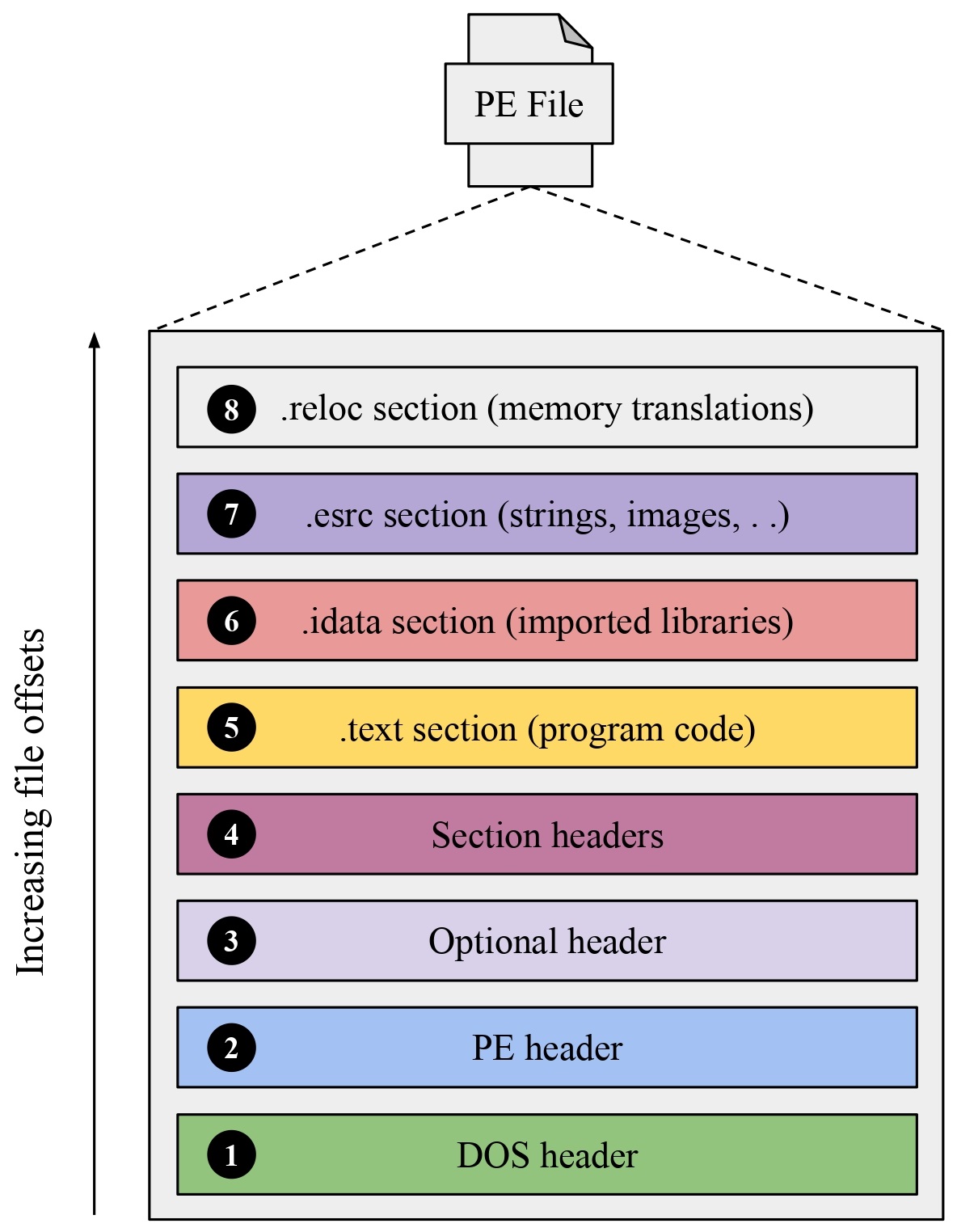}
        \caption{PE File Architecture}
        \label{fig:pefilearch}
    \end{subfigure}
    \hfill
    \begin{subfigure}[b]{0.45\textwidth}
        \centering
        \includegraphics[angle=0,origin=c, width=5.25cm]{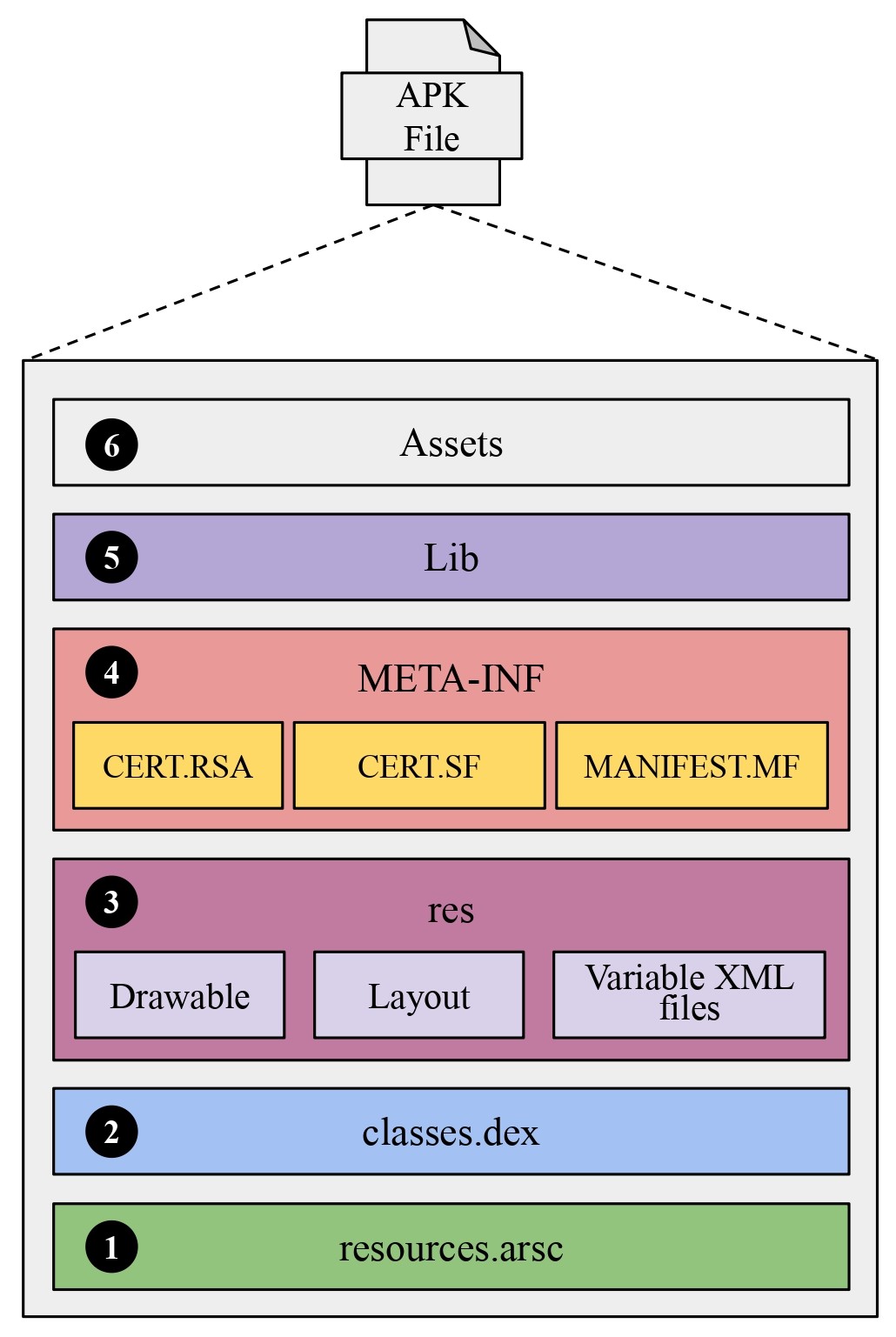}
        \caption{APK File Architecture}
        \label{fig:apkarch}
    \end{subfigure}
    \caption{PE and APK File Architecture}
    \label{fig:exearch}
\end{figure}

In the process of reverse engineering a Portable Executable (PE) file, one critical step is identifying the external functions it relies on, commonly referred to as API calls, which are essential for understanding the program’s behavior and interactions with the operating system. To achieve this, analysts must focus on the 
the PE file structure, shown in Figure~\ref{fig:pefilearch}, has different parts called sections. These sections are listed below from the start to the end of the file (increasing file offsets).

\subsubsection*{DOS Header}
The DOS header is the first part of the PE file. It’s like an old piece from the MS-DOS days. It helps the file work on older systems and points to the newer PE header.

\subsubsection*{PE Header}
The PE header comes next. It’s the main header that tells the computer how to read the file. It has important details like the file type and where other sections start.

\subsubsection*{Optional Header}
The optional header is the third part. It’s not always needed, but it has extra info like the size of the program and what system it runs on.

\subsubsection*{Section Header}
The section header is the fourth part. It’s like a map that lists all the sections in the file, such as where they are and how big they are.

\subsubsection*{.text Section}
The .text section is the fifth part. This is where the program’s code lives—the instructions that the computer runs when you open the file.

\subsubsection*{.idata Section}
The .idata section is the sixth part, placed above the .text section and below the .esrc section, as shown in Figure~\ref{fig:pefilearch}. It’s also called the import directory. This section has a list called the Import Address Table (IAT), which shows all the outside tools (APIs) the program needs from other files called DLLs, like kernel32.dll or user32.dll. You can look at this section using tools like PE Explorer~\cite{PEExplorer}, IDA Pro~\cite{IDAPro}, or Ghidra~\cite{Ghidra}. For example, in IDA Pro, you can go to the "Imports" tab to see the DLL names and the APIs the program uses. Checking the .idata section helps you understand what the program depends on and what it does, making it a key part to study in the file.

\subsubsection*{.esrc Section}
The .esrc section is the seventh part. It holds extra stuff like text strings, images, and other resources the program might need to show or use.

\subsubsection*{.reloc Section}
The .reloc section is the last part, the eighth one. It helps the program work in different spots in the computer’s memory by storing memory translation info.

\subsection*{APK File Structure}

The APK file structure, shown in Figure~\ref{fig:apkarch}, has different parts called sections. These sections are listed below from the start to the end of the file.

\subsubsection*{resources.arsc}
The resources.arsc is the first part of the APK file. It’s like a storage box for all the app’s resources, such as text strings and settings, that the app needs to work.

\subsubsection*{classes.dex}
The classes.dex is the second part. This is where the app’s code lives. It has all the instructions that tell the Android device what the app should do. The classes.dex file contains the compiled Dalvik bytecode of the application, which includes the Java or Kotlin code turned into a format that can run on the Android Runtime (ART) or Dalvik Virtual Machine (DVM). This bytecode has the API calls the app makes to talk to the Android system, system libraries, or other third-party libraries. To see these API calls, you can use tools like APKTool, dex2jar, or IDA Pro. For example, with dex2jar, you can turn the classes.dex file into a JAR file, then use JD-GUI to change it back into Java code, showing the API calls. Or, in IDA Pro, you can load the APK, go to the classes.dex section, and look at the Dalvik bytecode to find the methods the app uses, helping you understand how the app works with other functions. So, the classes.dex section is the main place to check for understanding how the app uses APIs, giving important clues about what the app does and what it needs in the Android system.

\subsubsection*{res}
The res section is the third part. It holds resources like pictures, layouts, and other files the app uses. It has subfolders like:
\begin{itemize}
    \item Drawable: For images and icons.
    \item Layout: For how the app’s screens look.
    \item Variable XML files: For other settings the app needs.
\end{itemize}

\subsubsection*{META-INF}
The META-INF section is the fourth part. It has important files that keep the app safe and show it’s real. It includes:
\begin{itemize}
    \item CERT.RSA and CERT.SF: These are security files that prove the app is genuine.
    \item MANIFEST.MF: A file that lists all the other files in the APK and checks they haven’t been changed.
\end{itemize}

\subsubsection*{Lib}
The lib section is the fifth part. It has extra code libraries the app might need to work on different devices, like special code for certain types of Android hardware.

\subsubsection*{Assets}
The assets section is the last part, the sixth one. It stores raw files like fonts, sounds, or other data the app uses, which aren’t in the resources section.

\noindent\textbf{Advantages:} Static analysis is fast, does not require runtime resources, and can detect malicious intent in dormant code.\\  
\textbf{Limitations:} It struggles with obfuscated or packed malware, where system calls are hidden or dynamically resolved at runtime.\\  
\textbf{Example:} A static scan of a binary revealing repeated calls to \texttt{NtWriteFile} might suggest file encryption, a hallmark of ransomware, though confirmation requires further context.

\subsection{Dynamic System Call Analysis}
Dynamic analysis monitors system calls during program execution, capturing real-time behavior. This is achieved using tools like \texttt{strace} (Linux), Process Monitor (Windows)\cite{ProcessMonitor}, or Frida (Android)\cite{Frida}, which log system call sequences as the program runs. For example, a malware sample might invoke \texttt{socket} and \texttt{sendto} during execution, indicating data exfiltration.

\noindent\textbf{Advantages:} Dynamic analysis captures actual runtime behavior, bypassing obfuscation techniques like code packing.\\  
\textbf{Limitations:} It requires a controlled environment to avoid harm, may miss dormant malicious code (e.g., logic bombs), and can be evaded by malware detecting monitoring tools.\\  
\textbf{Example:} Tracing a trojan on Linux might show \texttt{fork} followed by \texttt{execve}, revealing a process-spawning attack, observable only during execution.

\subsection{Sandboxing to Extract System Calls}
Sandboxing involves executing a program in an isolated, virtualized environment to safely observe its system calls and behavior. Sandboxes simulate an operating system, logging interactions like file access, network traffic, and memory usage. They are widely used in malware analysis to combine the benefits of dynamic analysis with safety and scalability.
\begin{figure}[h!]%
    \centering
    {
        {\includegraphics[angle=0,origin=c, width=12cm]{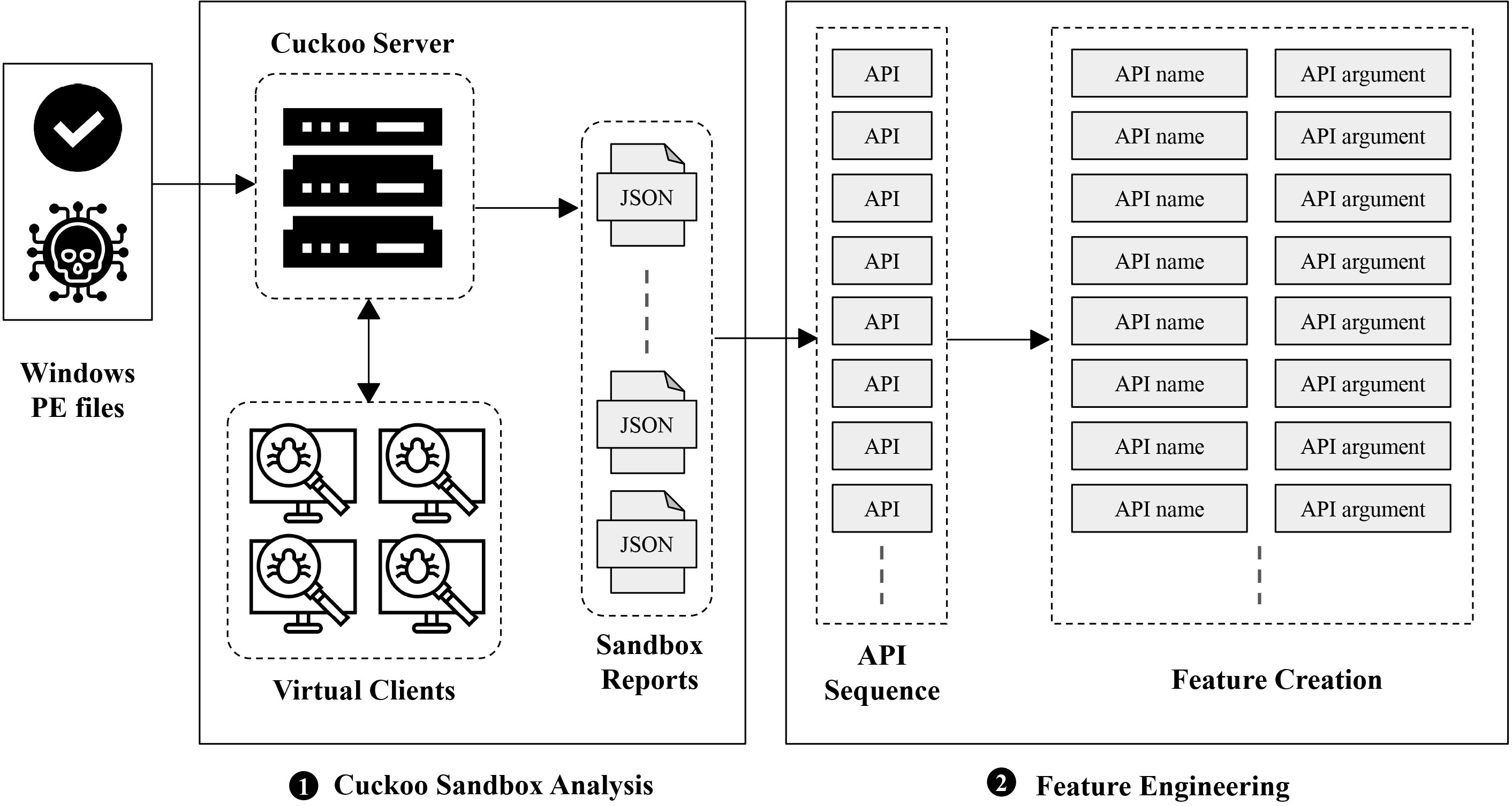} }
    }%
    \centering
    \caption{Cuckoo Sandbox Analysis and feature engineering }
    \label{fig:cuckoofeatureeng}%
\end{figure}

A sandbox\index{sandbox} is a security mechanism used in software development and testing, as well as in cybersecurity, to isolate running programs or processes from the rest of the system as shown in phase one of Figure~\ref{fig:cuckoofeatureeng}. The concept is inspired by children's sandbox play areas, where kids can play freely without affecting the surrounding environment. In the context of software, a sandbox provides a controlled environment where untrusted or potentially harmful code can be executed without posing a risk to the underlying system.

\subsubsection{Working Principle of a Sandbox}

\begin{enumerate}
    \item \textbf{Isolation}: Sandboxes work by isolating the execution of applications or processes from the rest of the system. This isolation prevents any malicious or unintended actions performed by the sandboxed code from affecting the host system.
    \item \textbf{Resource Control}: Sandboxes typically control the resources available to the processes running within them. This includes limiting access to files, network resources, memory, and CPU usage. By restricting access to critical resources, sandboxes can mitigate potential damage caused by malicious code.
    \item \textbf{Monitoring and Analysis}: Sandboxes often include monitoring and analysis capabilities to observe the behaviour of the code running within them. This may involve logging system calls, network activity, file operations, and other interactions with the environment. Analysing this data can help detect and analyse suspicious or malicious behaviour.
    \item \textbf{Dynamic Analysis}: Sandboxes employ dynamic analysis techniques to evaluate the behaviour of code in real-time. This allows them to detect and respond to threats as they occur, such as identifying attempts to exploit vulnerabilities or execute malicious actions.
    \item \textbf{Containment}: If a sandbox detects malicious behaviour, it can take measures to contain the threat, such as terminating the offending process or reverting any changes made to the system.
\end{enumerate}

\subsubsection{Open-Source Sandboxes}
Open-source sandboxes are freely available tools developed by the community, offering flexibility and transparency. Examples include:
\begin{itemize}
    \item \textbf{Cuckoo Sandbox:} A popular open-source solution supporting Windows, Linux, and Android. It logs system calls, API calls, and network activity, generating detailed reports. For instance, Cuckoo\cite{CuckooSandbox} might detect a malware calling \texttt{NtCreateFile} to drop a payload.
    \item \textbf{Limon\cite{Limon}:} A Linux-specific sandbox built on \texttt{strace} and \texttt{ptrace}, focusing on native system call tracing. It excels at lightweight analysis of Linux malware.
\end{itemize}
\textbf{Advantages:} Cost-free, customizable, and widely supported by the research community.  \\
\textbf{Limitations:} Limited support, reliance on volunteer updates, and less polished user interfaces compared to commercial options.
\begin{figure}[h!]%
    \centering
    {
        {\includegraphics[angle=0,origin=c, width=16cm]{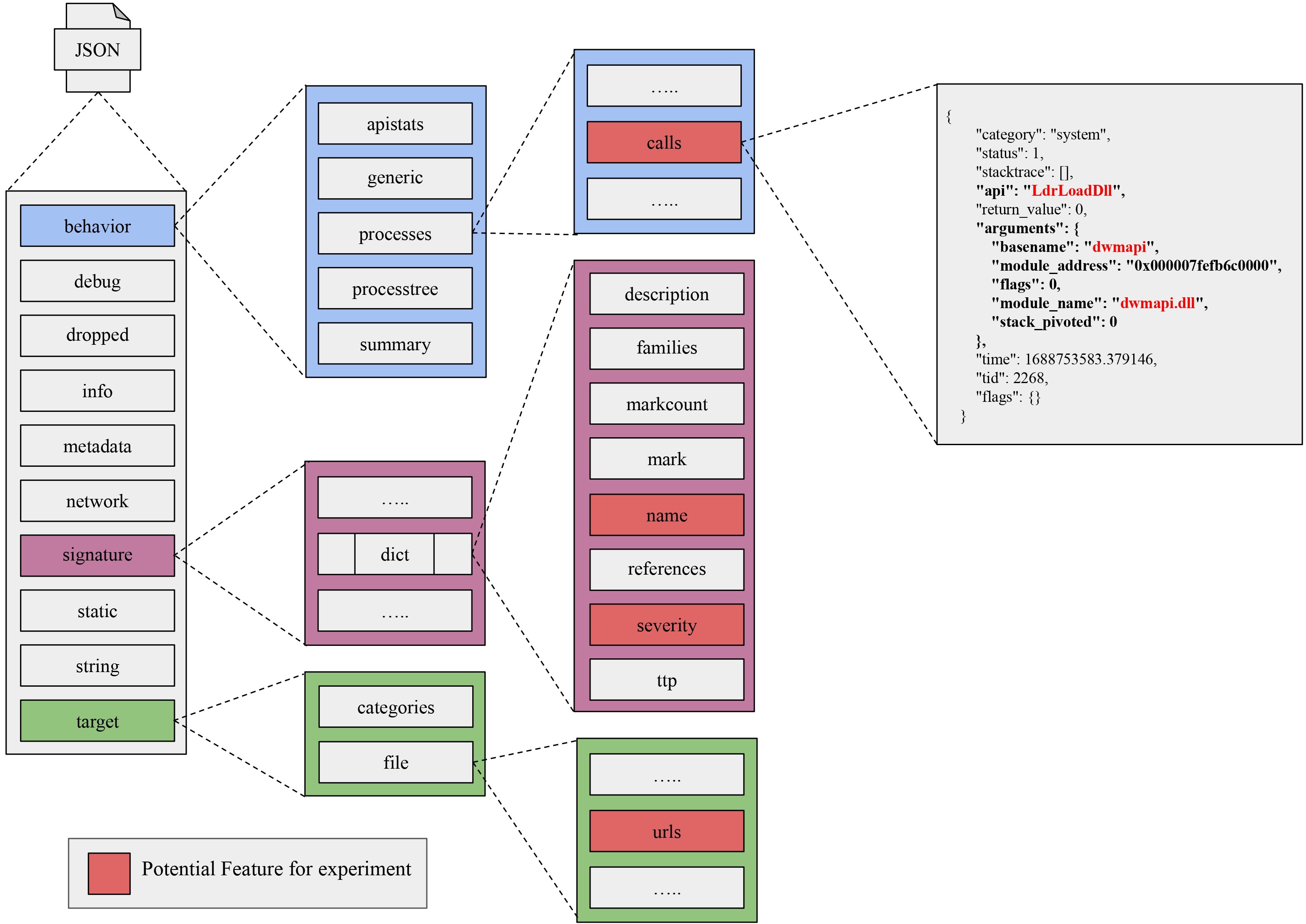} }
    % {\includegraphics[width=10cm]{Chap06/cuckooJsonAnalysis.jpg} }
    }%
    \centering
    %\captionsetup{format=plain}
    \caption{JSON Report Feature Selection after Cuckoo Analysis \cite{gonddeep} }
    \label{fig:cuckooana}%
\end{figure}
\subsubsection{License/Enterprise Sandboxes}
License or enterprise-grade sandboxes are commercial solutions designed for scalability, advanced features, and professional support. Examples include:
\begin{itemize}
    \item \textbf{FireEye Malware Analysis\cite{FireEye}:} An enterprise tool offering deep system call tracing across multiple platforms, with integrated threat intelligence. It might identify a zero-day exploit via \texttt{NtAllocateVirtualMemory}.
    \item \textbf{Joe Sandbox\cite{JoeSandbox}:} A license sandbox supporting Windows, Linux, and Android, with detailed behavioral analysis and cloud-based reporting. It excels at detecting sophisticated malware evading basic sandboxes.
\end{itemize}
\textbf{Advantages:} Robust feature sets, regular updates, and customer support, making them suitable for large-scale deployments. \\ 
\textbf{Limitations:} High costs and potential vendor lock-in, which may deter smaller organizations or researchers.

\subsubsection{Comparison and Use Cases}
Open-source sandboxes like Cuckoo are ideal for academic research or small-scale analysis due to their free accessibility and adaptability. In contrast, enterprise solutions like FireEye are better suited for organizations requiring real-time threat detection and integration with broader security ecosystems. Both approaches extract system calls effectively, but license tools often provide deeper insights into obfuscated or evasive malware through proprietary heuristics.

\textbf{Example:} A sandboxed Android app calling \texttt{bind} and \texttt{send} might be flagged as spyware. Cuckoo could detect this in a basic setup, while Joe Sandbox might also identify obfuscated API calls triggering the behavior.
\subsection{Discussion}
Static analysis is great for quickly scanning software before it runs, but it often falls short when malware uses obfuscation to hide its behavior during execution. In contrast, dynamic analysis and sandboxing can uncover threats by observing how programs behave in real time, though these methods tend to be more resource-heavy and can sometimes be sidestepped by sophisticated malware. Sandboxing, in particular, strikes a good balance—it offers a safe but realistic environment where suspicious code can be tested without putting actual systems at risk. Whether to use an open-source or commercial sandbox depends largely on your budget, the size of your operation, and how advanced your detection needs are. Each option brings its own strengths to the table, helping to improve system call-based malware detection in different ways.

\section{Applications and Use Cases}\label{sec:appli}
There are several ways to effectively utilize a system or API Call for malware detection and classification using deep learning techniques, various feature extraction approaches can be employed depending on the model architecture. First, feature vectors extracted from malware binaries or logs can be passed as arguments to an API that interacts with a deep learning model such as a feedforward neural network or LSTM. This structured numerical representation captures key behavioral or static properties of the malware. For a similarity-based approach using n-grams, the system processes n-gram sequences (e.g., byte-level or opcode sequences), converts them into feature vectors based on frequency or embedding similarity, and uses these vectors as inputs to the API call. Finally, for CNN-based analysis, the API accepts inputs where malware binaries have been converted into grayscale images (e.g., byte-to-pixel mapping) and processed as image data. The API call with this image argument allows the CNN to extract spatial patterns typically associated with malware families or variants. These approaches allow the malware detection system to support multiple deep learning backends while abstracting the input format through structured API calls.  Below we present some of the real-world applicatons used as use cases.

\subsection{API call with argument Feature Vector n-gram using similarity approach}
This subsection discusses the methodology for using n-grams of API call sequences, including both API names and their arguments, to create feature vectors for malware classification using a similarity-based approach, as detailed in the paper ``NLP-Driven Malware Classification: A Jaccard Similarity Approach'' \cite{gond2024nlp}. The approach utilizes Natural Language Processing (NLP) techniques combined with Jaccard similarity to capture behavioral patterns in malware samples, enabling effective classification across various malware types.

The process begins with dynamic analysis of malware samples in a Cuckoo Sandbox environment, where behavioral reports are generated in JSON format. These reports are segmented into four components: API category, API name, API argument, and API return. For feature extraction, the API name and argument are combined to form n-grams (unigrams, bigrams, and trigrams), with arguments appended to API names using underscores (e.g., \texttt{LdrLoadDll\_urlmon\_urlmon.dll} for a unigram). This representation captures the contextual relationship between API calls and their parameters, reflecting malware behavior more comprehensively than API names alone.

\begin{figure}[h!]%
    \centering
    {{\includegraphics[width=7cm]{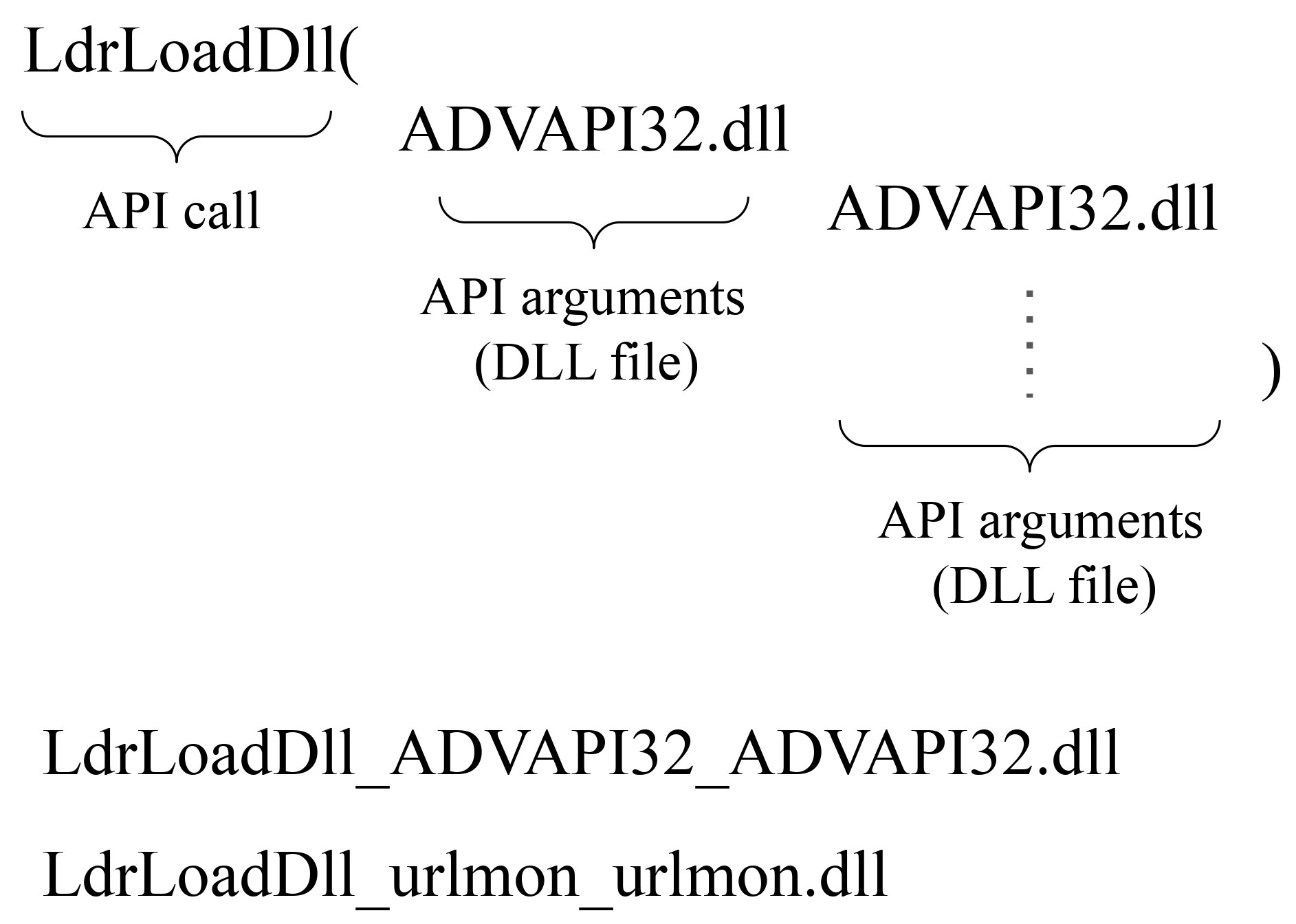} }}%
    \centering
    \caption{Feature creation from API Sequence \cite{gond2024deep}}
    \label{fig:feature}%
\end{figure}

After analyzing the PE files and generating JSON data, we segmented the behavioral processes of function calls into four components: API name, API return, API argument, and API categories as shown in Figure \ref{fig:cuckoofeatureeng}. From these four segments, we concatenated the API name and argument together, as illustrated in Figure \ref{fig:feature}. In this figure, the API call is mentioned first, followed by all DLL files invoked by the API in the same sequence of calls. This approach enables a more comprehensive understanding of the function call behavior within the analyzed PE files, facilitating a clearer insight into the runtime interactions and dependencies of the software components.

\subsection*{Measuring Text Similarity with Jaccard Similarity and TF-IDF}\label{sec:jstf}

\subsubsection*{Preprocess the Data}
We start by preprocessing the input data and reference data, which are in the form of data frames. We convert each data frame into a single text string. This text string represents all the content within each data frame, which can include multiple rows and columns.
\subsubsection*{$n$-grams after Cuckoo Sandbox Analysis} The following are the examples of unigrams, bigrams and trigrams that we have used in this research project.
\begin{itemize}
    \item \textbf{Unigram: }\textit{NtAllocateVirtualMemory\_na}
    \item \textbf{Bigram: }\textit{NtAllocateVirtualMemory\_na,\\ LdrLoadDll\_ole32\_ole32.dll}
    \item \textbf{Trigram: }\textit{NtAllocateVirtualMemory\_na,\\ LdrLoadDll\_ole32\_ole32.dll\\ LdrGetProcedureAddress\_ole32\_OleUninitialize}
\end{itemize}

\subsubsection*{Vectorization with TF-IDF}
Term Frequency-Inverse Document Frequency (TF-IDF) is a numerical statistic that reflects how important a word is to a document in a collection or corpus. We use TF-IDF to convert the text strings into numerical vectors. Each word in the text strings becomes a feature, and the TF-IDF score of each word in each text string is calculated. The TF-IDF score reflects the importance of the word within the text string. Common words like "the" will have lower TF-IDF scores, while unique words will have higher scores. The TF-IDF vectors represent the content of the data in a way that can be used for similarity measurement.
\begin{figure}[h!]%
    \centering
    {
        {\includegraphics[angle=0,origin=c, width=8cm]{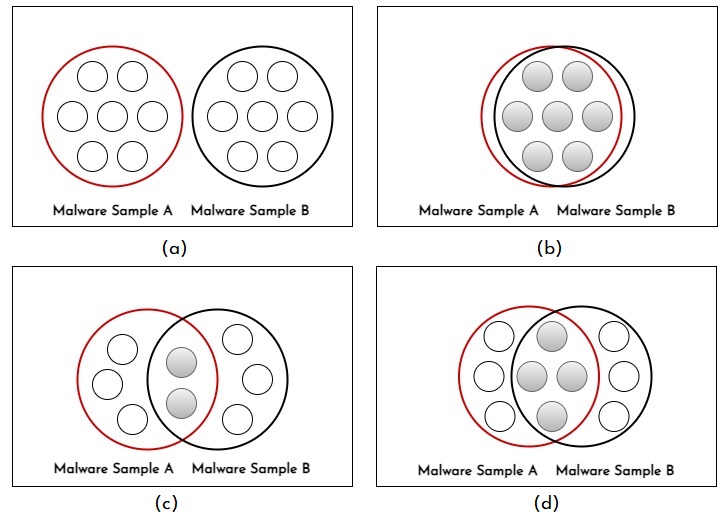} }
    }%
    \centering
    \caption{Similarity Based Approach \cite{gond2024nlp}}
    \label{fig:cuckooana}%
\end{figure}
\begin{figure}[h!]%
    \centering
    {
        {\includegraphics[angle=0,origin=c, width=16cm]{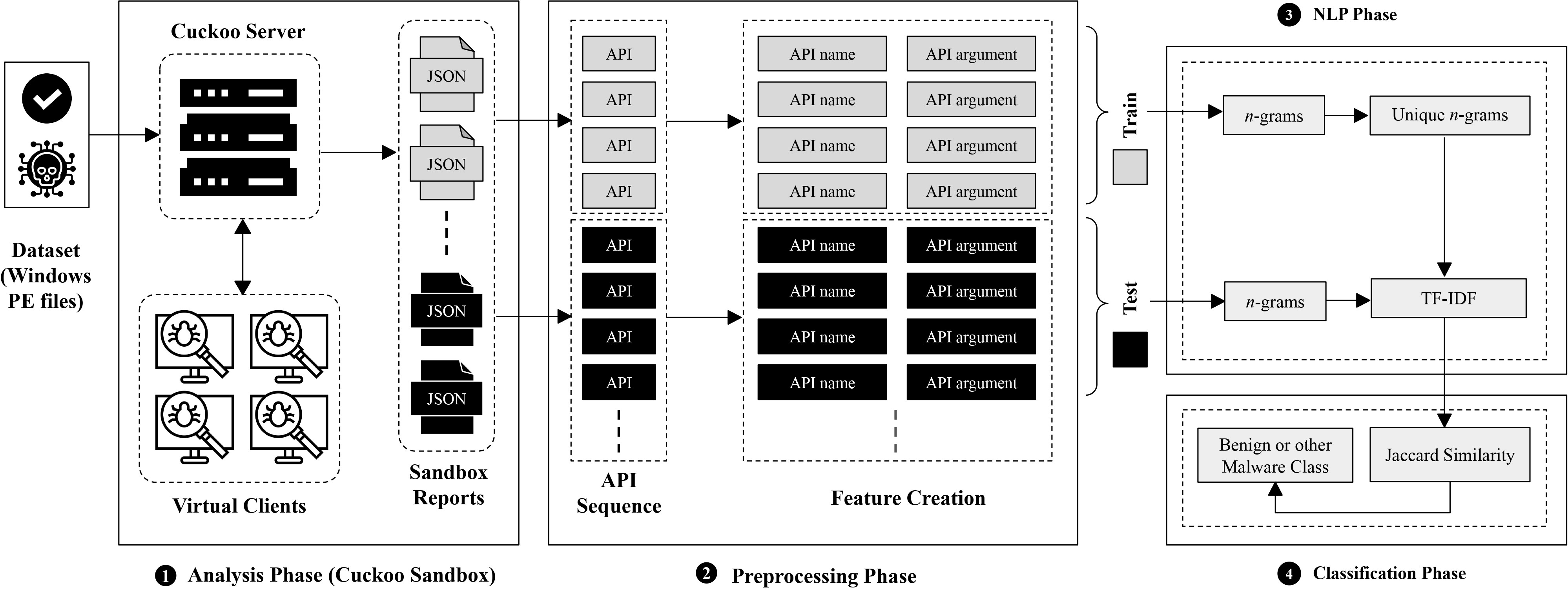} }
    }%
    \centering
    \caption{Malware Classification using NLP and Jaccard Similarity \cite{gond2024nlp}}
    \label{fig:nlparchjaccard}%
\end{figure}
\subsubsection*{Calculate Jaccard Similarity}
Jaccard similarity  is a measure used to compare the similarity and dissimilarity between two sets. It is defined as the size of the intersection of the sets divided by the size of the union of the sets. In other words, it calculates the ratio of the number of common elements to the total number of distinct elements in the sets. Mathematically, the Jaccard similarity (\( J(A, B) \)) between sets \( A \) and \( B \) is represented as: The formula for Jaccard similarity (J) between two sets A and B is given in Eq. \ref{eq:jaccard}.

\begin{equation}
    J(A, B) = \frac{|A \cap B|}{|A \cup B|}\label{eq:jaccard}
\end{equation}

In our case, the sets A and B represent the unique words (tokens) in two text strings.

The n-grams are processed to create a refined feature set by applying Term Frequency-Inverse Document Frequency (TF-IDF) vectorization. TF-IDF weights emphasize the importance of each n-gram relative to the corpus, enhancing the discriminative power of the features. Subsequently, Jaccard similarity is computed to quantify the similarity between n-gram sets of different malware samples. The Jaccard similarity metric, defined in Eq.~\ref{eq:jaccard}, measures the proportion of shared n-grams relative to the total unique n-grams, providing a robust indicator of behavioral overlap as shown in Figure~\ref{fig:nlparchjaccard}. A similarity matrix is constructed, where each element represents the Jaccard similarity between pairs of malware samples, facilitating clustering and classification.
\begin{figure}[ht!]
    \centering
    \begin{subfigure}[t]{0.32\textwidth}
        \centering
        \includegraphics[width=\textwidth]{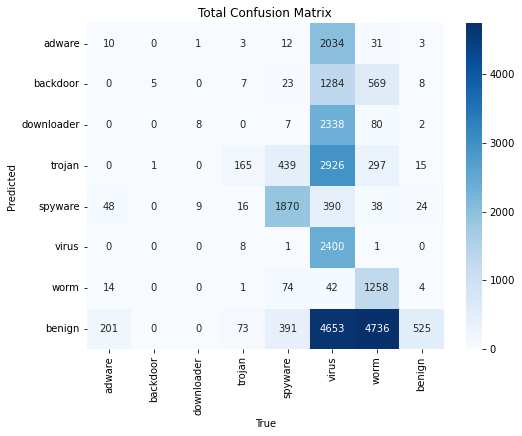}
        \caption{Confusion matrix for malware classification (unigrams)}
        \label{fig:uniconfm}
    \end{subfigure}
    \hfill
    \begin{subfigure}[t]{0.32\textwidth}
        \centering
        \includegraphics[width=\textwidth]{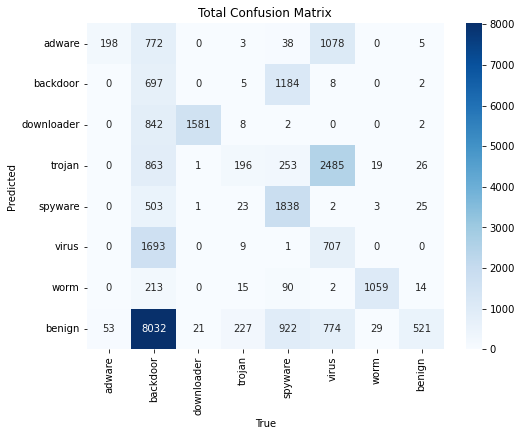}
        \caption{Confusion matrix for malware classification (bigrams)}
        \label{fig:biconfm}
    \end{subfigure}
    \hfill
    \begin{subfigure}[t]{0.32\textwidth}
        \centering
        \includegraphics[width=\textwidth]{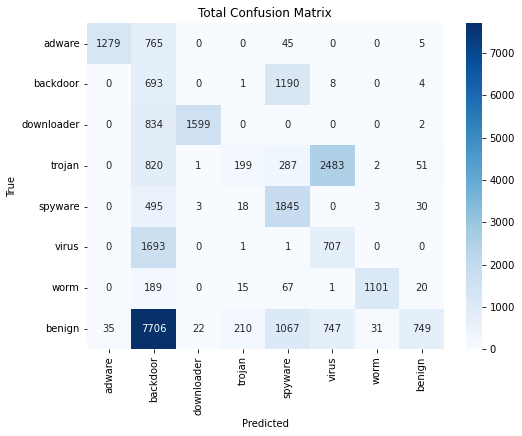}
        \caption{Confusion matrix for malware classification (trigrams)}
        \label{fig:triconfm}
    \end{subfigure}
    \caption{Confusion matrices for malware classification \cite{gond2024nlp}}
    \label{fig:confmats}
\end{figure}
% \begin{figure}[ht!]%
%     \centering
%     {{\includegraphics[width=8cm]{images/unigram_confusion_matrix.png} }}%
%     \centering
%     \caption{Confusion matrix for malware classification (unigrams)}
%     \label{fig:uniconfm}%
% \end{figure}

% \begin{figure}[htp!]%
%     \centering
%     {{\includegraphics[width=8cm]{images/bigram_confusion_matrix.png} }}%
%     \centering
%     \caption{Confusion matrix for malware classification (bigrams)}
%     \label{fig:biconfm}%
% \end{figure}

% \begin{figure}[ht!]%
%     \centering
%     {{\includegraphics[width=8cm]{images/trigram_confusion_matrix.png} }}%
%     \centering
%     \caption{Confusion matrix for malware classification (trigrams)}
%     \label{fig:triconfm}%
% \end{figure}

\begin{table}[ht!]
\centering
\caption{Datasets Used for Malware Analysis Using Jaccard Similarity \cite{gond2024nlp}}
\label{tab:maldata}
\begin{tabular}{l c c c r}
\toprule
\textbf{S.No} & \textbf{Type} & \textbf{Train Sample} & \textbf{Test Sample} & \textbf{Total Sample} \\
\midrule
1 & Adware & 2094 & 2500 & \textbf{4594} \\
2 & Worm & 1390 & 2500 & \textbf{3890} \\
3 & Virus & 2410 & 2500 & \textbf{4910} \\
4 & Backdoor & 1896 & 2500 & \textbf{4396} \\
5 & Spyware & 2395 & 2500 & \textbf{4895} \\
6 & Benign & 11000 & 10579 & \textbf{21579} \\
7 & Trojan & 3838 & 2500 & \textbf{6338} \\
8 & Downloader & 2435 & 2500 & \textbf{4935} \\
\midrule
\multicolumn{1}{l}{\textbf{Total}} & & \textbf{27458} & \textbf{28079} & \textbf{55537} \\
\bottomrule
\end{tabular}
\end{table}

Experimental results demonstrate that higher-order n-grams (e.g., trigrams) yield improved classification performance, with worms achieving an accuracy of 98.78\% and an \( F_1 \)-score of 87.04\% using 3-gram features as shown in Figure~\ref{fig:confmats}. The approach excels in identifying nuanced patterns, particularly for malware types like downloaders and spyware, due to the detailed representation of API call sequences. The use of a large dataset (55,537 samples from VirusShare\cite{VirusShare} ) ensures generalizability, while the similarity-based method offers computational efficiency compared to traditional techniques.

\subsection{API call with argument Feature Vector using deep learning }
This subsection elaborates on the use of n-grams of API call sequences, including API names and arguments, as feature vectors for malware classification using deep learning techniques, as presented in the paper ``A Deep Learning Framework for Malware Classification using NLP Techniques''\cite{gond2024deep}. The approach harnesses the power of deep learning models, specifically Artificial Neural Networks (ANN), Convolutional Neural Networks (CNN), and Recurrent Neural Networks (RNN), to capture complex patterns in malware behavior.

The methodology starts with dynamic analysis in a Cuckoo Sandbox, producing JSON-formatted behavioral reports for malware samples. These reports are divided into API category, API name, API argument, and API return components. The API name and argument are combined to form n-grams (e.g., \texttt{NtAllocateVirtualMemory\_na} for a bigram), which serve as feature vectors. These n-grams are vectorized using Term Frequency (TF) calculations, where,
\begin{equation}
     TF(t, d) = \frac{f_{t, d}}{\sum_{t^{\prime} \in d} f_{t^{\prime}, d}} 
\end{equation}
to quantify the importance of each n-gram within a sample. A hybrid feature selection process reduces the feature set from 5,578,098 to 88,975 (approximately 1.6\% of the original), eliminating redundant or irrelevant features to enhance model efficiency.
\begin{figure}[h!]%
    \centering
    {
        {\includegraphics[angle=0,origin=c, width=15cm]{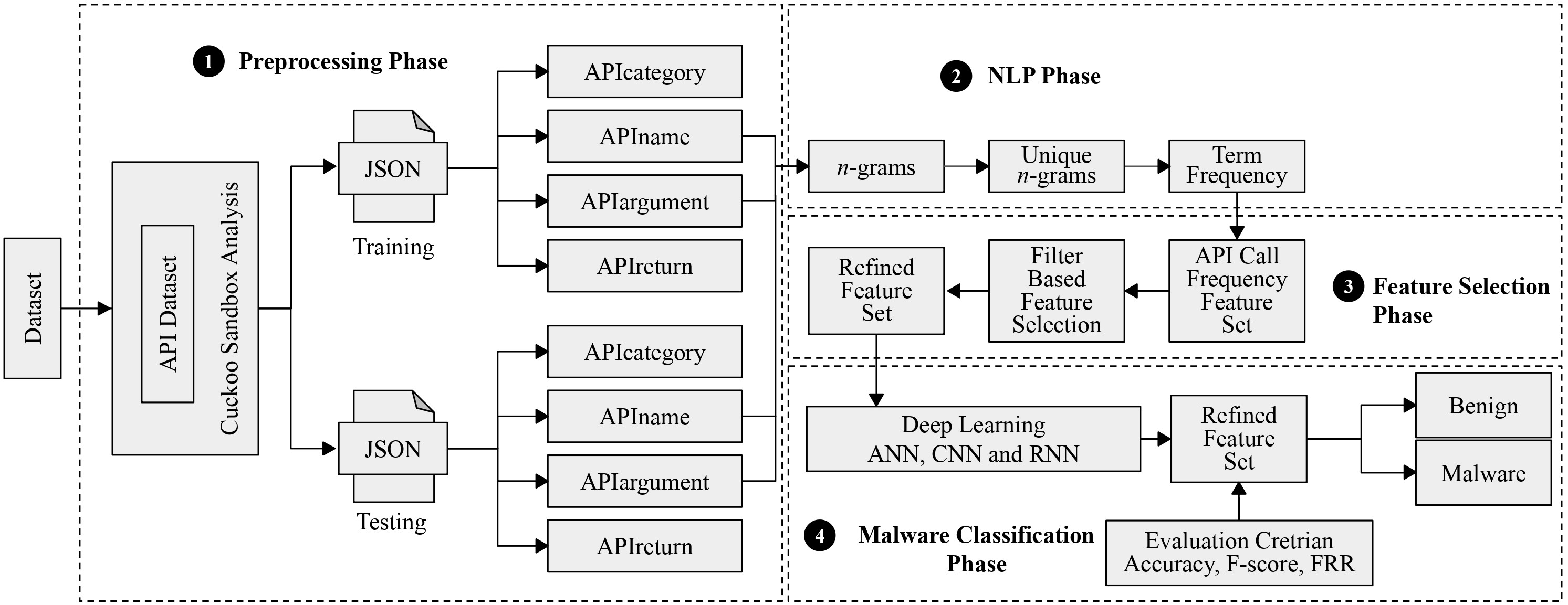} }
    }%
    \centering
    \caption{Malware Classification using Deep Learning Techniques \cite{gond2024deep}}
    \label{fig:nlparchdl}%
\end{figure}
Three deep learning models are employed for classification:
\begin{enumerate}
    \item \textbf{ANN Model}: Comprises of multiple dense layers with tanh activation, dropout layers (rate 0.4), and a softmax output layer for eight classes. It achieves a validation accuracy of 97.35\% after 100 epochs.
    \item \textbf{CNN Model}: Features convolutional layers with ReLU activation, max-pooling, and dense layers, yielding the highest performance with a validation accuracy of 98.53\% and a loss of 0.0161 at 100 epochs.
    \item \textbf{RNN Model}: Uses a SimpleRNN layer with ReLU activation, but struggles with a constant accuracy of 39.16\%, indicating unsuitability for this task.
\end{enumerate}
\begin{table}[ht!]
\centering
\caption{Datasets Used for Malware Analysis Using Deep Learning \cite{gond2024deep}}
\label{tab:mal_data}
\begin{tabular}{l c c c r}
\toprule
\textbf{S.No} & \textbf{Type} & \textbf{Test Sample} & \textbf{Train Sample} & \textbf{Total Sample} \\
\midrule
1 & Adware & 406 & 1580 & \textbf{1986} \\
2 & Spyware & 190 & 756 & \textbf{946} \\
3 & Trojan & 695 & 2873 & \textbf{3568} \\
4 & Backdoor & 123 & 551 & \textbf{674} \\
5 & Virus & 500 & 1892 & \textbf{2392} \\
6 & Worm & 277 & 1080 & \textbf{1357} \\
7 & Benign & 1724 & 6910 & \textbf{8634} \\
8 & Downloader & 495 & 2002 & \textbf{2497} \\

\midrule
\multicolumn{1}{l}{\textbf{Total}} & & \textbf{4410} & \textbf{17644} & \textbf{22054} \\
\bottomrule
\end{tabular}
\end{table}
% \begin{table}[ht!]
% \caption{Datasets used}
% \begin{center}
% \begin{tabular}{|c|c|c|c|c|}
% \hline
% %\multicolumn{5}{|c|}{\textbf{Malware Types}} \\
% %\hline\hline
% \textbf{S.No} & \textbf{{Types}}& \textbf{{Test Sample}}& \textbf{{Train Sample}}& \textbf{{Total Sample}} \\
% \hline\hline
% % 1& Adware$^{\mathrm{a}}$&2500 & 2500 \\
% % \hline
% 1& Adware &406 & 1580 &\textbf{1986} \\
% \hline
% 2& Backdoor &123 & 551  &\textbf{674}\\
% \hline
% 3& Downloader &495 & 2002 &\textbf{2497}\\
% \hline
% 4& Spyware &190 & 756 &\textbf{946}\\
% \hline
% 5& Trojan &695 & 2873 &\textbf{3568}\\
% \hline
% 6& Worm &277 & 1080 &\textbf{1357}\\
% \hline
% 7& Virus &500 & 1892 &\textbf{2392}\\
% \hline
% 8& Benign &1724 & 6910 &\textbf{8634}\\
% \hline\hline
% &\multicolumn{1}{|c|}{\textbf{Total}} & \textbf{4410} &\textbf{17644} &\textbf{22054}\\
% \cline{2-5} 
% \hline
% \end{tabular}
% \label{tab:mal_data}
% \end{center}
% \end{table}

\begin{table}[h!]
\centering
\caption{Summary of Loss and Accuracy for Different Epochs of ANN, CNN, and RNN}
\label{tab:combined_epochs}
\begin{tabular}{l c c c c c c}
\toprule
\textbf{Model} & \textbf{No.} & \textbf{Epoch} & \textbf{Loss} & \textbf{Val Loss} & \textbf{Accuracy} & \textbf{Val Accuracy} \\
\midrule
\multirow{6}{*}{ANN} 
& 1 & 1   & 1.1005 & 0.5596 & 0.6485 & 0.8203 \\
& 2 & 20  & 0.1959 & 0.1678 & 0.9461 & 0.9658 \\
& 3 & 40  & 0.1524 & 0.1490 & 0.9577 & 0.9694 \\
& 4 & 60  & 0.1096 & 0.1400 & 0.9730 & 0.9703 \\
& 5 & 80  & 0.0869 & 0.1258 & 0.9780 & 0.9723 \\
& 6 & 100 & 0.0761 & 0.1226 & 0.9799 & 0.9735 \\
\midrule
\multirow{6}{*}{CNN} 
& 1 & 1   & 7.8877 & 4.6905 & 0.7281 & 0.8645 \\
& 2 & 20  & 0.1254 & 0.1722 & 0.9691 & 0.9764 \\
& 3 & 40  & 0.0505 & 0.1292 & 0.9875 & 0.9850 \\
& 4 & 60  & 0.0248 & 0.1570 & 0.9927 & 0.9866 \\
& 5 & 80  & 0.0312 & 0.1081 & 0.9937 & 0.9859 \\
& 6 & 100 & 0.0161 & 0.1846 & 0.9950 & 0.9853 \\
\midrule
\multirow{6}{*}{RNN} 
& 1 & 1   & 1.9359 & 1.8120 & 0.3754 & 0.3908 \\
& 2 & 20  & 1.7850 & 1.7761 & 0.3916 & 0.3908 \\
& 3 & 40  & 1.8376 & 1.8367 & 0.3916 & 0.3908 \\
& 4 & 60  & 1.7867 & 1.7881 & 0.3916 & 0.3908 \\
& 5 & 80  & 1.7376 & 1.7967 & 0.3916 & 0.3908 \\
& 6 & 100 & 1.6867 & 1.8181 & 0.3916 & 0.3908 \\
\bottomrule
\end{tabular}
\end{table}

% \begin{figure}[ht!]
%     \centering
%     \begin{subfigure}[t]{0.32\textwidth}
%         \centering
%         \includegraphics[width=\textwidth]{images/CNN.png}
%         \caption{Loss and Accuracy measures for CNN model}
%         \label{fig:CNN}
%     \end{subfigure}
%     \hfill
%     \begin{subfigure}[t]{0.32\textwidth}
%         \centering
%         \includegraphics[width=\textwidth]{images/ANN.png}
%         \caption{Loss and Accuracy measures for ANN model}
%         \label{fig:ann}
%     \end{subfigure}
%     \hfill
%     \begin{subfigure}[t]{0.32\textwidth}
%         \centering
%         \includegraphics[width=\textwidth]{images/RNN.png}
%         \caption{Loss and Accuracy measures for RNN model}
%         \label{fig:rnn}
%     \end{subfigure}
%     \caption{Loss vs Accuracy measures for Deep learning model \cite{gond2024deep}}
%     \label{fig:dlimage}
% \end{figure}
The dataset, consisting of 22,054 samples as shown in Table \ref{tab:mal_data}, from VirusShare \cite{bishwajit_prasad_gond_2025}, is split into training (17,644) and testing (4,410) sets. The CNN model outperforms others, leveraging its ability to detect local patterns in n-gram sequences. The use of API arguments in n-grams enhances the model’s ability to differentiate malware types, achieving a classification accuracy of 99.5\% for the CNN model as shown in Table \ref{tab:combined_epochs}. This approach demonstrates robustness against evolving malware threats by capturing intricate behavioral patterns.

\subsection{API call with argument converted into Graph for malware detection and classification using GNN}
\begin{figure}[h!]%
    \centering
    {
        {\includegraphics[angle=0,origin=c, width=13cm]{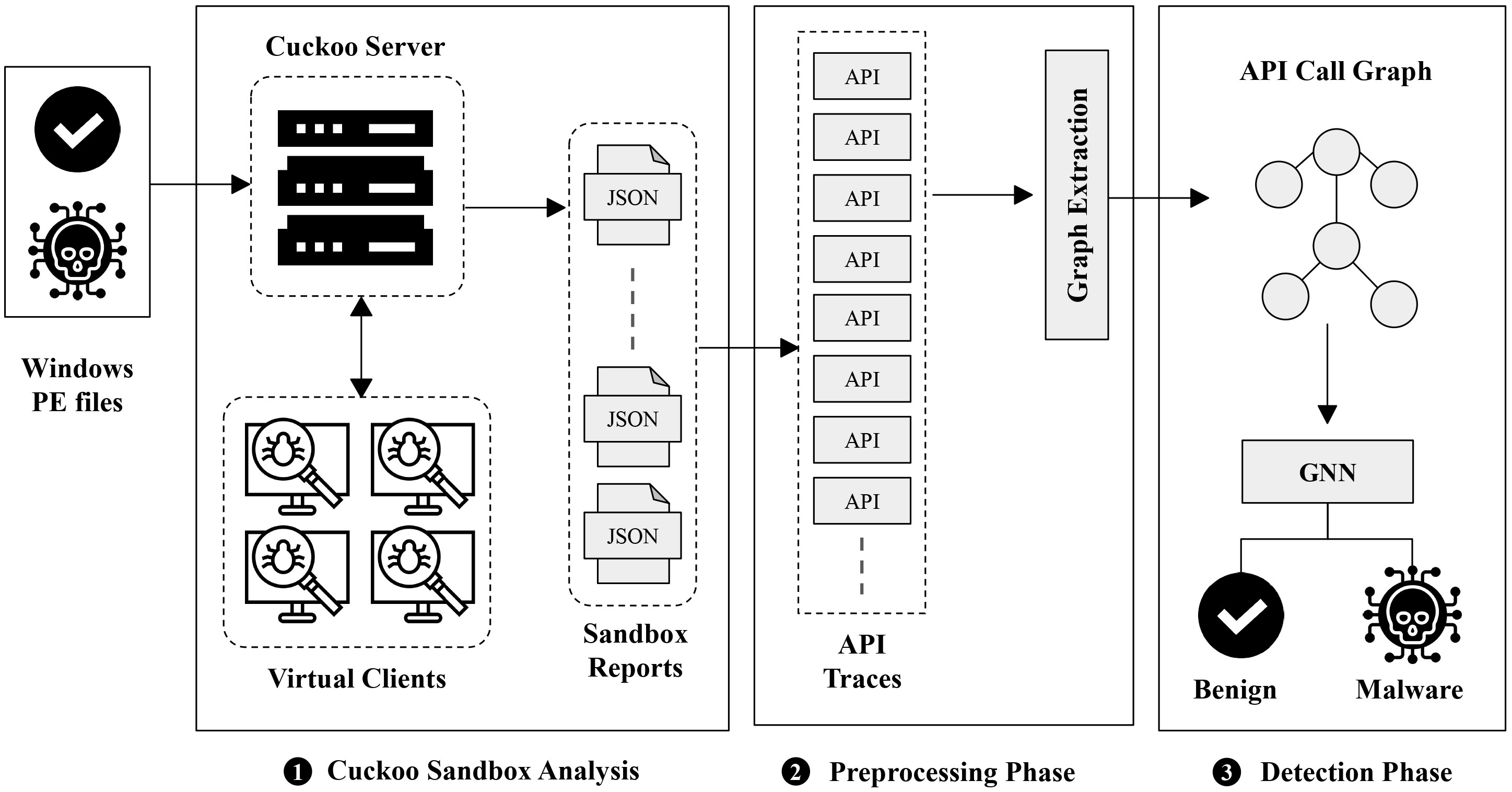} }
    }%
    \centering
    \caption{Malware Classification using Graph Neural Network \cite{rajneekant2024malware}}
    \label{fig:gn}%
\end{figure}

This subsection elaborates on the use of n-grams of API Call Embedding and Graph Neural Networks, as presented in the paper  ``Malware Detector and Classifier Using API Call Embedding and Graph Neural Networks" \cite{rajneekant2024malware} addresses the growing threat of malware, particularly on Windows OS. The proposed framework leverages API call sequences, including arguments, to enhance malware detection and classification. It employs Levenshtein distance for API call embedding to capture nuanced differences in arguments, addressing limitations of prior methods like Jaccard similarity. A graph is constructed from these embeddings, and a Graph Neural Network (GNN) model, incorporating linear transformations, graph convolution, and matrix factorization convolution layers, is used to analyze structural patterns as shown in Figure~\ref{fig:gn}.

\begin{table}[h]
\centering
\caption{GNN based malware detector performance parameters}
\label{tab:gnn_performance}
\begin{tabular}{l c}
\toprule
\textbf{Parameters} & \textbf{Value (\%)} \\
\midrule
Recall & 100.00 \\
F1-score & 99.83 \\
Accuracy & 99.70 \\
Precision & 99.66 \\
MCC & 99.59 \\
\bottomrule
\end{tabular}
\end{table}

The model was tested on a dataset of 41,483 samples from VirusShare\cite{VirusShare}, including various malware families and benign samples, achieving a detection accuracy of 99.70\%, an F1-score of 99.83\%, and a Matthews Correlation Coefficient (MCC) of 99.59\% as shown in Table \ref{tab:gnn_performance}. For classification, it attained 87.08\% accuracy and a 74.39\% MCC, with strong performance on families like downloader and ransomware but lower accuracy for trojans due to behavioral overlaps. Compared to state-of-the-art methods, the proposed model outperforms others in accuracy and F1-score, attributed to its robust feature extraction and GNN architecture. The study highlights the model’s efficiency, with a 20ms execution time per sample, comparable to commercial tools like VirusTotal. Future work aims to explore additional GNN architectures and improve explainability for detecting zero-day malware and enhancing cybersecurity.

\subsection{API call converted into Image for malware detection and classification using CNN}
Recent advancements in machine learning have inspired novel approaches to malware detection, notably the transformation of behavioral data into visual representations. One such technique involves converting sequences of API calls and their arguments into images, which are then analyzed using convolutional neural networks (CNNs) for detection and classification tasks.

This approach leverages the insight that API call patterns, along with their associated arguments (e.g., file names, registry keys, network addresses), encapsulate the behavioral signatures of malware. When encoded into image formats, these patterns can reveal latent structures more effectively captured by deep learning models. Nataraj et al.\cite{nataraj2011malware} were among the first to propose converting malware binaries into grayscale images for classification, demonstrating the potential of visual analysis in security contexts. Building on this idea, researchers have extended this method to dynamic behavioral data.

Tobiyama et al.\cite{tobiayama2016malware} developed a system that models malware behavior through long short-term memory (LSTM) networks and CNNs by visualizing API sequences as behavior graphs, which are then rendered into images for classification.

Ko et al.\cite{ko} introduces a novel malware detection method for Windows environments using a Convolutional Neural Network (CNN) that leverages 2-gram opcode frequency data transformed into images. Unlike traditional methods that rely on manual signature-based analysis or byte-to-image conversion, this approach extracts opcode sequences from disassembled executable files, computes their 2-gram frequencies, and converts these into spatial image data for CNN processing. The method employs Hierarchical Clustering to group highly correlated opcodes, positioning them closely in the image to capture semantic relationships. The CNN model, consisting of convolutional layers, max-pooling, batch normalization, and fully-connected layers, achieved a 91\% accuracy rate in detecting malware on a dataset of 10,000 malicious and 10,000 benign files. The study compares various clustering algorithms, finding Hierarchical Clustering most effective, and outperforms Multi-Layer Perceptron (MLP) models by 3–5\% in accuracy. The method’s use of opcode frequency images enhances semantic analysis, offering a scalable solution for automated malware detection. Future work aims to integrate this approach with other CNN-based frameworks to improve detection accuracy further.
These studies underline the effectiveness of converting API call arguments into image formats, enabling more accurate and scalable detection. Such transformations help abstract complex execution traces into forms suitable for deep learning, improving detection rates while reducing reliance on handcrafted features.

Shahnawaz et al.\cite{shahnawaz2025} introduces a novel image-based dynamic malware classification framework that transforms runtime API call arguments from Windows Portable Executable (PE) files into grayscale images. By utilizing Cuckoo Sandbox, the authors extract API behavior logs, focusing on API names and their arguments during execution. These are then converted into structured feature vectors, normalized, reshaped into 128×128 matrices, and enhanced with techniques like Gaussian blur, CLAHE, and Sobel edge detection. The resulting images are visualized using the magma colormap and used to train a custom Convolutional Neural Network (CNN). The CNN architecture comprises three convolutional and pooling layers, a dense layer, and dropout regularization, ending in a softmax classifier for multi-class prediction. The model is trained on a dataset of 22,056 samples across eight classes, achieving high accuracy (up to 100\%) and robustness, especially against obfuscated or polymorphic malware. With an average classification accuracy of 98.36\%, the approach proves scalable, interpretable, and resistant to common evasion tactics. Future directions include incorporating Large Language Models (LLMs) for API semantics, cross-platform adaptability, model compression for IoT use cases, and sequential modeling using LSTMs or Transformers.
\begin{figure}[h!]%
    \centering
    {
        {\includegraphics[angle=0,origin=c, width=13cm]{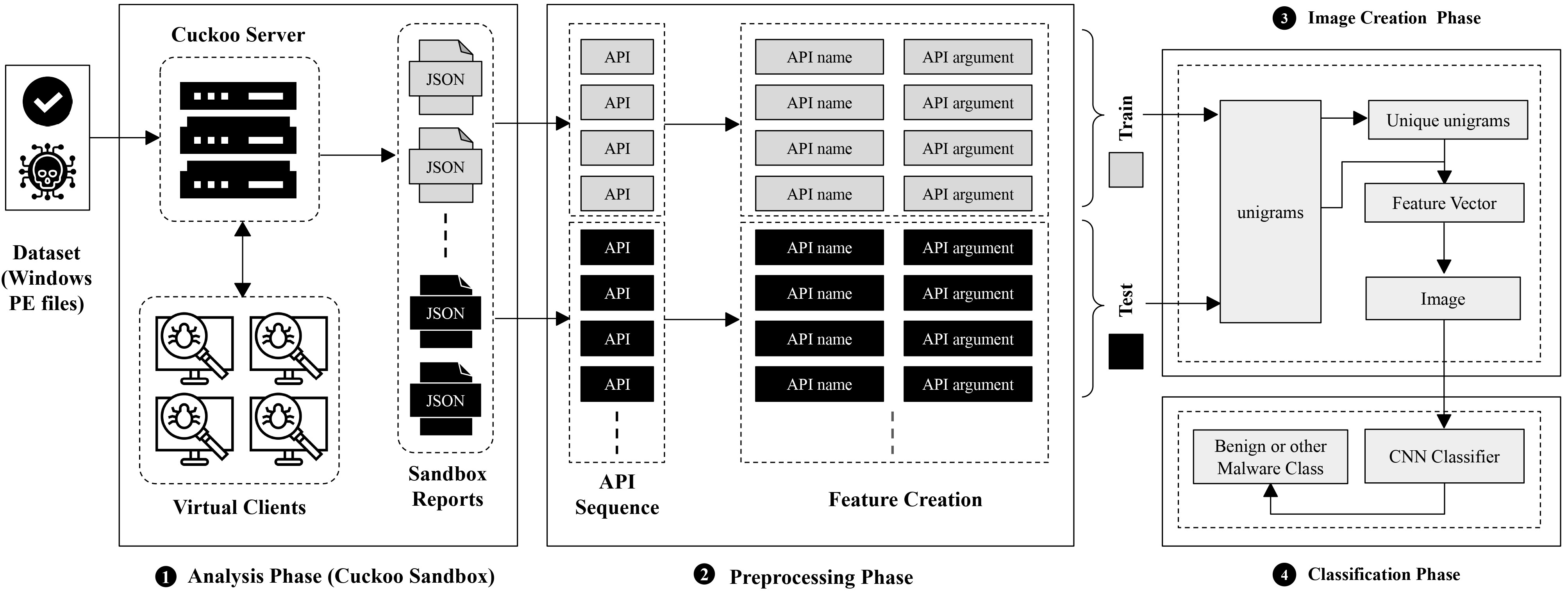} }
    }%
    \centering
    \caption{Malware Classification using Convolutional Neural Network \cite{shahnawaz2025}}
    \label{fig:gn}%
\end{figure}
\begin{figure}[h!]%
    \centering
    {
        {\includegraphics[angle=0,origin=c, width=13cm]{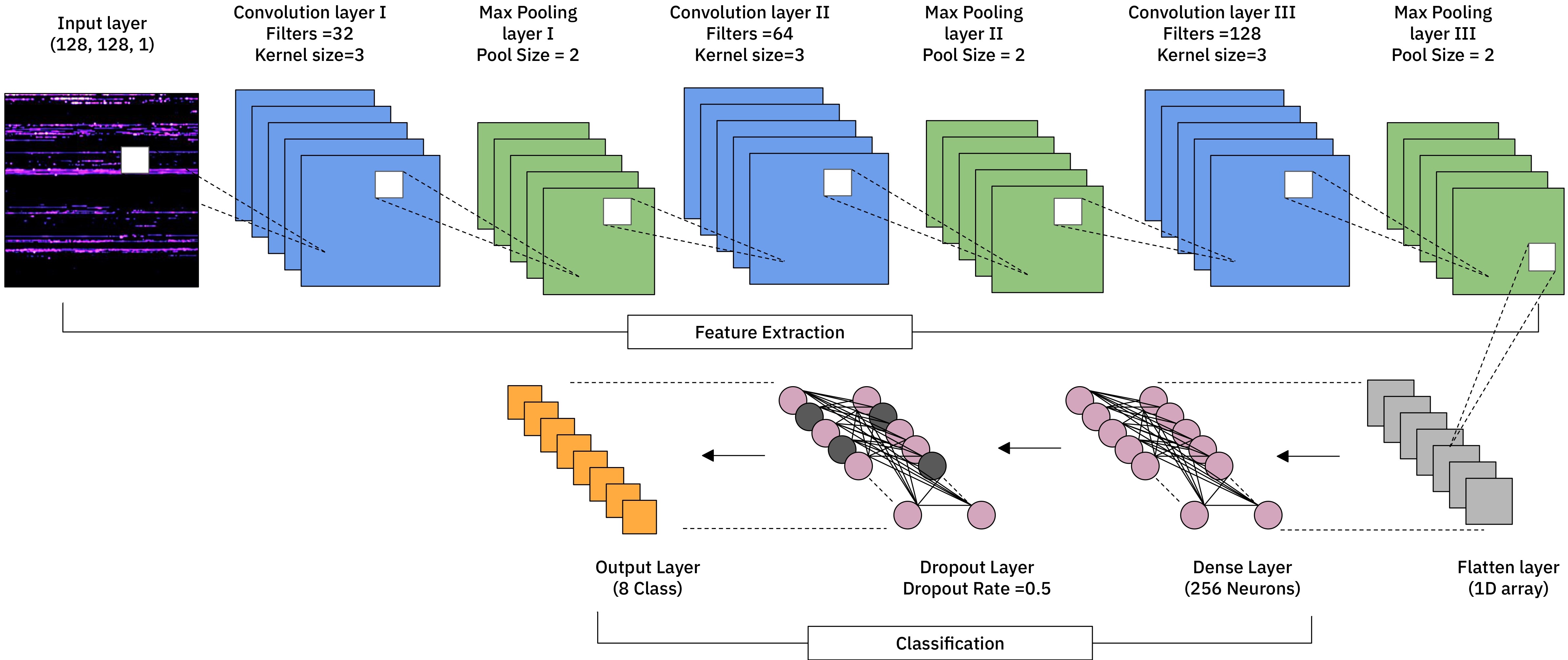} }
    }%
    \centering
    \caption{Detailed Convolutional Neural Network \cite{shahnawaz2025}}
    \label{fig:gn}%
\end{figure}

\begin{table}[h!]
\centering
\begin{tabular}{@{}lcccc@{}}
\toprule
\textbf{Malware Type} & \textbf{Accuracy} & \textbf{F1 Score} & \textbf{Recall} & \textbf{Precision} \\
\midrule
Adware    & 98.00\% & 87.50\% & 78.60\% & 98.70\% \\
Backdoor  & 99.20\% & 86.70\% & 84.40\% & 89.10\% \\
Benign    & 96.00\% & 95.10\% & 98.70\% & 91.80\% \\
Downloader& 100.00\% & 100.00\% & 100.00\% & 100.00\% \\
Spyware   & 99.20\% & 89.90\% & 84.70\% & 95.80\% \\
Trojan    & 97.60\% & 92.30\% & 89.80\% & 95.00\% \\
Virus     & 98.80\% & 94.60\% & 93.10\% & 96.10\% \\
Worm      & 98.10\% & 85.80\% & 91.10\% & 81.00\% \\
\bottomrule
\end{tabular}
\caption{Performance metrics of the proposed CNN-based malware classification model \cite{shahnawaz2025}}
\label{tab:performance}
\end{table}

\subsection{Discussion}
The similarity-based approach for malware classification, utilizing n-grams of API call sequences with arguments and Jaccard similarity, offers a computationally efficient method to capture behavioral patterns. As experimental results on a large dataset of 55,537 samples demonstrate high accuracy, particularly for worms (98.78\%) and spyware, due to the nuanced representation of API interactions. This method’s strength lies in its simplicity and ability to generalize across diverse malware types, though it may struggle with highly obfuscated samples due to reliance on surface-level similarities.

In contrast, the deep learning approach, which also uses n-grams of API calls and arguments, employs advanced models like CNNs, ANNs, and RNNs to uncover complex behavioral patterns, achieving superior performance with a CNN model accuracy of 99.5\% on a 22,054 sample dataset for classification and GNN model accuracy of 99.70\% on a 41,483 sample dataset for Detection. By combining API names and arguments into n-grams and applying TF-based feature selection to reduce the feature set to 88,975, this method enhances efficiency while capturing intricate relationships. The CNN and GNN models excels in detecting and capturing local patterns within n-gram sequences, making it robust against evolving malware threats. However, its computational complexity and reliance on large labeled datasets may limit scalability compared to the similarity-based approach, particularly in resource-constrained environments.

\section{Conclusion and Future Scope}\label{sec:future} 
Detecting and classifying malware through system call or APi Call analysis has become a powerful approach for spotting harmful software and program by looking at how programs interact with an operating system. Using static analysis can help catch threats quickly before a program even runs. However, this method often struggles with more advanced malware that uses techniques like obfuscation to hide its true nature. On the other hand, dynamic analysis where the software is run in a controlled environment or sandbox, it can expose malicious behavior in real time. While this gives a more complete picture, it also requires a secure setup to avoid any unintended consequences.

The above Use cases have brought impressive results using deep learning techniques. For example, researchers have used n-gram models, graph neural networks (GNNs), and even visual representations of API call sequences to improve detection. Convolutional neural networks (CNNs) have reached classification accuracies as high as 99.5\%, while GNNs have pushed detection rates to around 99.7\%. Open-source tools like Cuckoo Sandbox offer a low-cost, customizable way for researchers and smaller teams to experiment with these ideas. Meanwhile, commercial solutions such as FireEye provide large organizations with scalable, more sophisticated systems that integrate well into broader cybersecurity strategies.

Still, this field faces real challenges. Malware developers continue to find ways to hide from detection tools, and running dynamic analysis remains resource intensive. To address these issues, future research should focus on making detection methods more resistant to tricks like polymorphism and zero-day exploits. One promising direction is building hybrid models that combine both static and dynamic techniques, and use of LLM based Classifers finetuned with Malware API call Sequence that will offer a more well-rounded defense. Another key area is improving the transparency and explainbility of complex AI models such as GNN, CNN and etc, cybersecurity experts can better understand how and why a system flags certain behaviors.

% \bibliographystyle{unsrt}
% \bibliography{reference}

\end{document}